# Superconductivity in Pd-intercalated charge-density-wave rare earth poly-tellurides $R_E Te_n$

J. B. He[†,‡], P. P. Wang[†], H. X. Yang[†], Y. J. Long[‡], L. X. Zhao[‡], C. Ma[†], D. M. Wang[‡], X. C. Shangguan[‡], Z. A. Ren[†], J. Q. Li[†], and G. F. Chen[†,‡]

[†]Institute of Physics and Beijing National Laboratory for Condensed Matter Physics, Chinese Academy of Sciences, Beijing 100190, China

[‡]Department of Physics, Renmin University of China, Beijing 100872, China

The interplay between magnetism and superconductivity is one of the dominant themes in the study of unconventional superconductors, such as high-$T_c$ cuprates, iron pnictides and heavy fermions[1-3]. In such systems, the same *d*- or *f*-electrons tend to form magnetically ordered states and participate in building up a high density of states at the Fermi level, which is responsible for the superconductivity. Charge-density-wave (CDW) is another fascinating collective quantum phenomenon in some low dimensional materials, like the prototypical transition-metal poly-chalcogenides, in which CDW instability is frequently found to accompany with superconducting transition at low temperatures[4]. Remarkably, similar to the antiferromagnetic superconductors, superconductivity can also be achieved upon suppression of CDW order via chemical doping or applied pressure in 1*T*-TiSe$_2$ (ref. 5, 6). However, in these CDW superconductors, the two ground states are believed to occur in different parts of Fermi surface (FS) sheets, derived mainly from chalcogen *p*-states and transition metal *d*-states, respectively. The origin of superconductivity and its interplay with CDW instability has not yet been unambiguously determined. Here we report on the discovery of bulk superconductivity in Pd-intercalated CDW $R_E$Te$_n$ ($R_E$=rare earth; n=2.5, 3) compounds, which belong to a large family of rare-earth poly-chalcogenides with CDW instability usually developing in the planar square nets of tellurium at remarkably high transition temperature and the electronic properties are also dominated by chalcogen *p*-orbitals. Our study demonstrates that the intercalation of palladium leads to the suppression of the CDW order and the emergence of the superconductivity. Our finding could provide an ideal model system for comprehensive studies of the interplay between CDW and superconductivity.

Charge density waves (CDWs) are periodic modulations of the conduction-electron-density in solids. They often present in low dimensional electronic systems[4,7]. Usually, the CDW removes electrons from the Fermi level, and thus precludes a superconducting state[8-10]. However, in a variety of layered transition-metal dichalcogenides (TMDs), chain-type transition-metal trichalcogenides, and some organic compounds, CDW order coexists with superconductivity at low temperature[4,7,11]. The interplay between the two correlated electronic states has become a topic of central interest in condensed matter physics[12-14]. Recently, superconductivity has also been found to arise upon suppression of CDW order in $1T$-TiSe$_2$ via chemical doping[5] or the application of pressure[6]. The CDW-superconductivity phase diagram is quite similar to those of heavy fermions, high-$T_c$ cuprates and pnictides, in which the magnetic order is suppressed towards zero temperature through chemical doping or pressure[1-3]. The superconductivity appears close to magnetic phase boundaries suggesting that magnetic fluctuations are responsible for the pairing mechanism[15-17]. However, in the case of $1T$-TiSe$_2$, in spite of extensive theoretical and experimental efforts, the mechanism at the origin of the CDW and superconductivity is still unclear[18-23]. It is particularly interesting to explore new systems with superconductivity closely proximity to the CDW instability, and to systematically study the interplay between CDW and superconductivity.

$R_E$Te$_n$ ($R_E$=Y, La-Sm, Gd-Tm; n=2, 2.5, 3), a large family of rare-earth poly-tellurides, has attracted much attention in recent years[24-31]. Its particular interest lies in the correlation of the unique crystal structure with the charge-density wave formation. Remarkably, the rare earth tellurides show large anisotropic gap and very stable surface[24,25], which provide us an exceptional opportunity to study CDW states using angle resolved photoemission spectroscopy (ARPES). $R_E$Te$_n$ takes a layered, weakly orthorhombic crystal structure, in which square-planar Te sheets are sandwiched by corrugated double layers of $R_E$Te[26,31]. $R_E$Te$_2$ contains a single layer of Te sheet, $R_E$Te$_3$ contains double layers of Te sheets connecting via Van der Waals force, while for n=2.5, i.e.

$R_{E2}Te_5$, the crystal structure can be viewed as alternatively stacking $R_ETe_3$ and $R_ETe_2$ slabs along the b-axis , which contains both single and double Te square sheets. The corrugated $R_E$Te-slab is insulating and responsible for magnetism in the case of magnetic rare earth ion $R_E$, while the square Te-layer forms a two-dimensional conduction band[31-34]. It is very different from the case of layered TMD's, in which the triangular planes of transition metal ions dominate the electronic properties[12,35]. In these rare earth poly-tellurides, the incommensurate/commensurate CDWs are also localized on the planar square nets of tellurium with a remarkably high transition temperature, as confirmed by the observation of superstructure above the room temperature[28, 29, 36, 37], where the CDW is demonstrated to be driven by the imperfect nesting of Fermi surfaces (FS) [24, 33]. The observed metallic resistivity behavior in $R_ETe_n$ (n=2.5, 3) confirmed further that the FS remains in a certain momentum region of the Brillouin zone even in the CDW state[30, 32, 38]. The large anisotropy of the electrical conductivity demonstrates the strong two-dimensional nature due to the weak hybridization between the Te and the $R_E$Te layer[30, 38]. Especially, $R_{E2}Te_5$ may represents a rare example to show a hybrid CDW distortion with two independent modulation vectors from two different Te nets[28, 29].

For $R_ETe_3$ and $R_{E2}Te_5$, due to the weak Van der Waals bonding between the adjacent square Te-sheets, the crystals have potential to allow interlayer insertion of guest metal atoms or organic molecules. In fact, a series of new compounds $AMR_ETe_4$ has been successfully synthesized by insertion of $AM$Te layer in $R_ETe_3$ (A=K, Na; M=Cu, Ag; $R_E$=La, Ce, Eu)[39, 40]. In this case, there is no electron transferring between the [$AM$Te] and [$R_ETe_3$] slabs. An evident change in the CDW character has been attributed to the alternation occurring in the interlayer coupling between adjacent Te nets of $R_ETe_3$ slabs upon insertion of the $AM$Te layer[41, 42]. In this work, we report the discovery of bulk superconductivity in $R_ETe_n$ induced by Pd-intercalation (i.e., $Pd_xR_ETe_n$). The superconductivity is observed not only for the nonmagnetic rare earth elements Y, La, but also for the magnetic rare earths, Pr, Sm, Gd, Tb, Dy, Ho, Er, Tm, in which antiferromagnetism arising from the local moments on the rare earth ions is found to coexist with superconductivity. In particular, we focus on two typical series, $Pd_xHoTe_3$ and $Pd_xY_2Te_5$. For $Pd_xHoTe_3$, with increasing the Pd content (x), the CDW transition is continuously suppressed, and the superconducting state appears at x=0.04. Within the range of solid solution, the superconducting transition temperature $T_c$ has a dome-like shape in the doping-temperature phase diagram with a maximum near x=0.08 ($T_c$=2.83 K). The CDW-superconductivity phase diagram is quite similar to that of high-$T_c$ cuprates or iron pnictides[1,2]. For $Pd_xY_2Te_5$, Pd-intercalation only occurs between the double layers of Te sheets, in which the CDW can be suppressed by continuous Pd doping, and superconductivity emerges with a maximum $T_c$=3.05 K (x=0.08). TEM results confirmed that Pd-intercalation have little effect on the CDW state located on single Te square sheets. To our knowledge, this is the first example of superconductivity in an intercalated "rare earth" telluride. Our finding demonstrates the potential for exploring new superconductors in other rare-earth poly-chalcogenides, a large family of CDW hosts in which the conduction band is also predominantly formed from chalcogen *p*-orbitals. Furthermore, $Pd_xR_ETe_n$ offers an excellent opportunity to study the CDW to superconductivity transition through an easily controllable chemical parameter. Due to the same *p*-electrons are responsible for both superconductivity and CDW, it is very important to verify whether the origin of superconductivity in this family is the same as for TMD, or if other mechanisms are involved. Such systematic studies are believed to provide further insights into their possible underlying mechanisms.

Polycrystalline samples of $Pd_xR_ETe_n$ (0≤x≤0.3) were synthesized by solid-state reaction method. First, stoichiometric amount of rare earth elements $R_E$ (3N5), Te (5N), and Pd (3N) powders were sealed in thick walled quartz tube under vacuum. The quartz tube was heated to moderate temperature (500-700℃) slowly and held there for 50 hours, and then cooled to room temperature naturally. Second, the resulting product was reground, pressed into a pellet, sealed in an evacuated quartz tube and reheated for a further 50 hours. After repeating this process several times, homogeneous samples were obtained. The X-ray diffraction (XRD) pattern of the resulting samples was taken on a PANalytical diffractometer using Cu $K_\alpha$ radiation at room temperature. Magnetization measurements were carried out in MPMS SQUID VSM (Quantum Design) with zero-field-cooled (ZFC) and/or field-cooled (FC) modes. Transport and heat capacity properties were measured in PPMS system (Quantum Design). Microstructure analysis and in situ Transmission electron microscopy (TEM) observations were performed on a Tecnai F20 transmission electron microscope, equipped with low-temperature holders.

Figure 1a shows the highly two-dimensional crystal structure of $R_E Te_n$ family ($R_E$=Y, La-Sm, Gd-Tm): double sheets of nearly square planar tellurium (Te) are separated by a corrugated $R_E$-Te slab. Figure 1b and 1c show the powder X-ray diffraction patterns of $HoTe_3$ and $Y_2Te_5$ and Pd-doped samples with formula $Pd_xR_ETe_n$ (where n=2.5, 3; x= 0-0.25), which can be well indexed using orthorhombic crystal structure ($NdTe_n$-type) with the space group $Cmcm$[31]. Detailed analysis of the X-ray data demonstrates an evident increase of lattice constant b with the the increase of Pd doping content (x), as shown in Fig. 1d, indicating successful intercalation of the Pd into the weakly bonded double Te layers.

Figure 2a shows the temperature dependent resistivity $\rho(T)$ measured for polycrystalline samples of $Pd_xHoTe_3$. Consistent with previous work[38], the resistivity of $HoTe_3$ shows a clear, sharp jump at approximately 285 K, a broad maximum centered at about 260 K, and a good metallic behavior at low temperature. This feature has been attributed to the partial gapping of the Fermi surface at the CDW transition[43]. On doping with increasing amounts of Pd, the anomaly becomes broader, reduces in size and moves to the lower temperature. The superconductivity emerges at x=0.04, while a less pronounced anomaly is still observed at around 120 K. The highest $T_c$= 2.83 K is obtained at x=0.08, which can be seen clearly from an expanded plot of the temperature-dependent resistivity curve, as shown in Fig. 2b. $T_c$ is found to drop slightly with further Pd-doping. Interestingly, $\rho(T)$ shows a semiconducting-like behavior before the superconducting transition. A similar behavior has also been observed in Pd doped 1$T$-TiSe$_2$ with a limited doping range[44], in contrast to Cu doping of 1$T$-TiSe$_2$, where increasing Cu content continuously improve the metallicity[5]. Further investigation is certainly needed to clarify this point.

Figure 2c shows the temperature dependence of electrical resistivity ρ($T$) measured for polycrystalline samples of Pd$_x$Y$_2$Te$_5$. One can see that, for pure Y$_2$Te$_5$, with decreasing temperature, ρ(T) increases rapidly around 300 K, followed by a broad hump centered at about 200 K, and then decreases slowly. Such an anomaly observed in the family of rare earth tritellurides $R_E$Te$_3$ is associated with the CDW transition, but $R_E$Te$_3$ exhibits a good metallic behavior at low temperature[45]. Actually, we cannot clarify the CDW transition upon the presented data. The appearance of the prominent broad peak in ρ(T) for Y$_2$Te$_5$ seems to have another reason. Earlier studies have shown that, in $R_{E2}$Te$_5$, the single layer Te sheets are tellurium deficient, while the double Te sheets have essentially perfect square symmetry[30]. So, in $R_{E2}$Te$_5$, the large resistivity and the broad maximum far below room temperature, might be described by a two-component scaling function composed of a semiconducting phase and metallic phase, as observed $T$-dependent resistivity in the family of $R_E$Te$_{2-\delta}$ and $R_E$Te$_3$ (ref. 32, 46). With Pd doping, the huge hump anomaly shifts to lower temperature and becomes less pronounced. The superconductivity emerges at x=0.04, and the highest $T_c$=3.05 K is obtained at x=0.08. At higher temperatures, for samples with 0⩽x⩽0.06, ρ($T$) displays slightly metal-like behavior down to 200 K.

The bulk superconductivity in Pd-doped $R_E$Te$_n$ is confirmed by dc magnetic susceptibility, as shown in Fig. 3a and 3b. With increasing Pd content in Pd$_x$HoTe$_3$, and Pd$_x$Y$_2$Te$_5$, both the superconducting transition temperature and the diamagnetic signal increase and reach the maximum at x=0.08; further increase in x results in suppression of the signal strength and $T_c$. The shielding volume fraction at 1.8 K is estimated to be 100% for optimal doping sample. Specific heat $C_p$ measurement carried out at $H=0$, and 1 T for $Pd_{0.08}Y_2Te_5$, is shown in Fig. 3c. The peak was suppressed with applied magnetic field, as expected for a superconducting transition. One can see that the specific jump near $T_c$ is very sharp with $H=0$, indicating the good quality of the sample. Below 10 K, the normal state specific heat $C(T)$ can be well fitted by a relation, $C_p=\gamma T+\beta T^3$, indicating that the specific heat $C_p$ is mainly contributed by electrons and phonons. The fit yields the electronic coefficient $\gamma= 2$ mJ/molK$^2$, and the Debye temperature $\Theta_D = 207$ K. Using the $\gamma= 2$ mJ/molK$^2$ and $T_c= 3$ K, we obtained $\Delta C_P/\gamma T_c = 1.63$, slightly larger than the valve of 1.43 within the BCS weak coupling approximation.

In Fig. 3d, we present the temperature dependence of magnetic susceptibility $\chi(T)$ for $Pd_{0.08}HoTe_3$ in the low temperature region with a magnetic field of 10 Oe. Near 3.5 K, $\chi$ tends to show a small peak, and followed by a clear diamagnetic drop corresponding to superconducting transition at around 2.8 K, which corresponds to the zero resistivity temperature. To get more information about the magnetic properties, we performed the dc magnetic susceptibility measured at the magnetic field of 0.1 T, as shown in the lower inset of Fig. 3d. Above 50 K, the data are well fitted with a simple Curie-Weiss law with an effective magnetic moment of 10.82 $\mu_B$, which is close to the magnetic moment of free $Ho^{3+}$ ion. The sharp drop below 3 K here is due to the AFM ordering, which is consistent with the specific heat data (the upper inset of Fig. 3d). All these results indicate that the superconductivity coexists with the AFM ordering of $R_E$ 4$f$ local moments due to the weak exchange interaction between the local rare earth 4$f$ moments and Te 5$p$ conduction electrons. Similar phenomena has been reported in other systems, such as $R_E Rh_4B_4$, $R_E Mo_6S_8$, and $R_E Ni_2B_2C$, in which the crystallographic site for the magnetic ions ($R_E$ ions) is well isolated from the conduction path [47-49].

As discussed above, for $Pd_{0.08}Y_2Te_5$ without the rare earth 4*f* electrons, the superconducting transition is clearly observable at $T_c \sim 3$ K. On the contrary, there is no specific heat discontinuity surrounding $T_c$ related to superconducting transition for $Pd_{0.08}HoTe_3$, as shown in the upper inset of Fig. 3d, due to the large background specific heat caused by both magnetic and crystal electric field (CEF) effects, which impedes the resolution of the superconducting jump[45]. Similar behavior has also been seen for the typical magnetic superconductor $HoNi_2B_2C$ (ref. 50). In particular, in this case, the superconducting transition temperature is very close to the AFM order ($T_N/T_c<1.2$), so it is not possible to extract the superconducting contribution from the strong background.

An important parameter to characterize superconductivity is the upper critical field $H_{c2}(0)$. We measured the temperature-dependent resistivity of the optimal doping samples, $Pd_{0.08}HoTe_3$, and $Pd_{0.08}Y_2Te_5$, under different magnetic fields, as shown in Fig. 4a and 4b. The corresponding upper critical field $H_{c2}$ versus $T_c$ is plotted in Fig. 4c and 4d, where $T_c$ is defined by a criterion of 50% of normal state resistivity. One can see that, with increasing the magnetic field, the transition temperature $T_c$ shifts to lower temperature and the transition width gradually becomes broader, similar to the high $T_c$ iron-based superconductors[2], suggesting the strong anisotropy of the critical field, as expected from the two-dimensional electronic structure[33]. The values of $-dH_{c2}/dT/T_c$ are -0.42 for $Pd_{0.08}HoTe_3$, and -0.62 for $Pd_{0.08}Y_2Te_5$, respectively. Using the Werthamer-Helfamd-Hohenberg formula[51], $H_{c2}(0)=-0.69(dH_{c2}/dT)T_c$, and taking $T_c=$ 2.8 K, and 3 K, the upper critical fields are estimated to be 0.81 T and 1.28 T, respectively. Figure 4e and 4f depict the *M(H)* versus *H* plot at 1.5-3 K measured by sweeping the magnetic field at a constant rate of 1 Oe/sec, which indicates that $Pd_xR_ETe_n$ is a type-II superconductor with a strong vortex pinning. The lower critical fields $H_{c1}$ were estimated to be 15 Oe for $Pd_{0.08}HoTe_3$ and 20 Oe for $Pd_{0.08}Y_2Te_5$, respectively, which are smaller than that of the copper-intercalated 1*T*-$TiSe_2$ with $T_c$ = 4.15 K (ref. 5).

Now let us take an overall view on the properties of other Pd-doped rare earth tritellurides. Figure 5 shows the low temperature dependent resistivity for $Pd_{0.08}R_ETe_3$ ($R_E$=La, Pr, Sm, Gd, Tb, Dy, Er and Tm) samples. One can see that, for the heavy rare earth tritellurides with the element changing from Gd, Tb to Tm, all compounds exhibit superconductivity with $T_c$~2-3 K, while for the light rare earth compounds $Pd_{0.08}LaTe_3$, $Pd_{0.08}PrTe_3$, $Pd_{0.08}SmTe_3$, they show superconductivity at relatively low temperatures ($T_c$<1 K). At present stage, resistivity measurements for Pd-doped $CeTe_3$ and $NdTe_3$ samples do not show superconductivity with the temperature down to 50 mK. Several possible reasons for the suppression of superconductivity in light rare earth compounds are considered here. Firstly, the suppression of superconductivity might arise from the magnetic pair breaking effect. For example, in $R_ENi_2B_2C$, the $T_c$ changes from 16.6 K in $(Lu,Y)Ni_2B_2C$ to 0 K for $TbNi_2B_2C$, in which the decrease of $T_c$ is roughly scaled by de Gennes factor DG= $(g-1)^2J(J+1)$, implying the dominance of the magnetic pair breaking effect in the systematic change of $T_c$ with the rare earth element[52]. However, in our case, the $T_c$ change between $Pd_xTmTe_3$ and $Pd_xGdTe_3$ is less than 1 K. Here the antiferromagnetism doesn't affect superconductivity dramatically regarding $T_c$, because the 4$f$ moments due to the $R_E$ sublattice is electronically isolated from the Te square nets, and hence no strong pair breaking occurs in the conducting square nets of Te. Secondly, the absence of superconductivity may be ascribed to the hybridization of the 4$f$ electrons with itinerant conduction electrons. Among of them, $CeTe_3$ shows a moderate heavy fermion behavior, and the occurrence of spin density wave (SDW) at low temperature ($T_{SDW}$ = 1.3 K) has been proposed according to the neutron diffraction and specific heat measurements[53]. Recent ARPES result also shows that, in $CeTe_3$, the 4$f^1$ peak moves closer to the Fermi level (-0.28 eV), which is actually different from other $R_ETe_3$ materials[43].

Lastly, although the Fermi surface is composed of states associated with the double Te layers in $R_ETe_3$, the change of the rare-earth element has been found to cause a remarkable difference in the periodic lattice distortion (PLD) nesting vector, as well as the transport and magnetic properties due to the effect of "chemical pressure"[32,36,38]. For example, the CDW transition temperatures reduce obviously from the light to heavy rare-earth elements[38]. A similar variation of the CDW properties could also be induced by applied pressure[54,55]. In addition, recent ARPES results have pointed out that the percentage of ungapped FS changes from 35% in CeTe$_3$ to 55% in DyTe$_3$, giving a strong $R_E$ dependence[40]. The change of lattice spacing causes the changed electronic structure compared to the cases of $R_E$ with small ionic radius, like Tm, in particular the reduced density of states at the Fermi level $N(E_F)$. We noticed that, for heavy rare earth tritellurides except for the Gd, Tb compounds, the second CDW transition $T_{CDW2}$ formed at the low temperature has been observed clearly in the temperature dependent resistivity for these parent compounds[35]. $T_{CDW2}$ was found to decease with the increasing of the lattice parameter, eventually vanishing for TbTe$_3$, which shows the opposite trend compared to the first CDW transition[35]. It seems that there is some connection between the chemical pressure induced second CDW transition and superconductivity. But the effect of lattice structure alone is not sufficient to explain the suppression of superconductivity for light rare earth tri-tellurides, since Pd$_x$LaTe$_3$ superconducts at $T_c$~0.6 K in spite of relatively large lattice constant ratio b/a. Therefore, it is reasonable to consider here that the suppression of superconductivity for light rare earth tellurides is ascribed to the combination of multiple effects.

Previous TEM and high resolution X-ray scattering measurements revealed that the rare-earth tritellurides $R_E$Te$_3$ compounds often incommensurate modulated structures arising from CDW, the modulation appears along c* axis direction with their wave vectors ranging from 0.2719 c* to 0.3028 c* for $R_E$ = La to Tm[56]. Moreover, it has been also demonstrated that $R_{E2}$Te$_5$ compounds contain two sets of structural modulations corresponding with two different CDW states in the single and double Te planes, respectively[29,54]; the visible modulation from the CDW state in the double Te planes adopts an incommensurate periodicity similar with what observed in $R_E$Te$_3$. On the other hand, the other modulations from a CDW state in the single Te planes have the commensurate nature. TEM investigations in $Gd_2Te_5$ revealed the commensurate structural modulations can be written as $q_1 = 5/12a^* + 1/12c^*$ and $q_2 = 1/12a^* + 5/12c^*$ (ref. 57).

In order to reveal the structural features in correlation with the CDW transitions and superconductivity in the rare-earth poly-tellurides, we have carried out a number of in-situ TEM observations with a number of well-characterized samples. For the parent phases $R_ETe_3$ and $R_{E2}Te_5$, similar structural modulations and microstructure features for CDW states have been observed as reported in previous literatures[26, 28, 57]. Moreover, our careful analysis suggests that the incommensurate modulation become evidently invisible in superconducting samples, this fact demonstrates that the CDW states formed in the double Te planes is efficiently suppressed in association with the appearances of superconductivity. Figures 6a and 6b show, respectively, the [010] zone-axis Selected Area Electron Diffraction (SAED) patterns taken at the temperature around 100 K for $TmTe_3$ and $Pd_{0.08}TmTe_3$. It is recognizable that notable incommensurate diffraction spots, as indicated in Fig. 6a, disappear completely in Fig. 6b which is obtained from a superconducting sample. Similar experimental results have been also obtained based on diffraction observations for the $Y_2Te_5$ CDW phase and $Pd_{0.08}Y_2Te_5$ superconducting phase, as clearly illustrated in the low-temperature SAED patterns of Fig.6c and 6d, respectively.

Figure 7 is the phase diagram showing the resistivity anomaly (open circles) and superconducting transition (closed circles) temperatures as a function of Pd content in $Pd_xHoTe_3$. The $T$-x electronic phase diagram is quite similar to that of $Cu_xTiSe_2$ (ref. 5), where the superconductivity emerges when the CDW order is destroyed by doping. However, one should be noted that, an important difference between the two is that, for Cu-intercalated $1T$-$TiSe_2$, the doped electrons enter the bottom of the Ti 3$d$-induced conduction band, which dominates the electronic properties[35, 58], while the Pd atom has a $4d^{10}5s^0$ configuration, it seems that no electron transfer between Pd and $R_ETe_n$. Furthermore, the Fermi surface of $R_ETe_n$ is derived mainly from the $p$-orbitals of the Te atoms, which are formed the double layers of nearly square-planar sheets;

and the contribution near $E_F$ from $R_E$ 4$f$ states is negligible[31-34]. Further theoretical and spectroscopic studies will be helpful to understand the underlying mechanism.

In summary, we have synthesized series of materials Pd$_x$$R_E$Te$_n$ by solid-state reaction. The parent compounds, $R_E$Te$_n$, develop CDW instability with a moderately high transition temperature. It is found that the CDW order could be suppressed continuously by intercalation of Pd, and induces bulk superconductivity for a limit range of Pd concentrations. Such a phase diagram is reminiscent of that in the well studied copper-intercalated 1$T$-TiSe$_2$, while the simplicity of the electronic structure of the Te planes in $R_E$Te$_n$ makes it a good candidate to carry out a complete study of the relation between the CDW and superconductivity.

## Acknowledgements

This work was supported by the Natural Science Foundation of China, and by the Ministry of Science and Technology of China.

**Figure Legends**

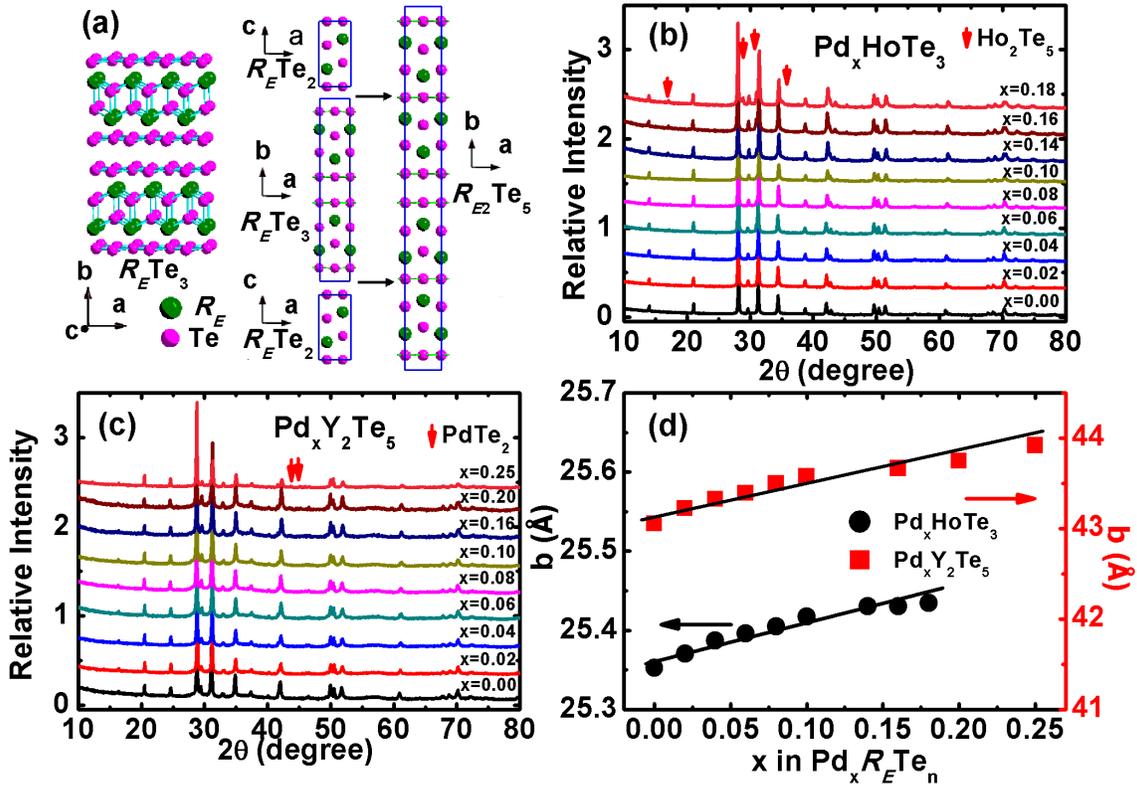

**Figure 1. Crystal structural analyses of Pd$_x$HoTe$_3$ and Pd$_x$Y$_2$Te$_5$. a**, Schematic crystal structures of $R_E$Te$_2$, $R_E$Te$_3$ and $R_{E2}$Te$_5$. The green and pink circles represent rare earth $R_E$ and Te ions, respectively. Note that the $R_E$Te-slab and Te-sheet are stacked along the b-axis. Pd ions are considered to be intercalated into the Van der Waals layer between double Te-sheets. **b, c**, Powder X-ray diffraction data collected for Pd$_x$HoTe$_3$ and Pd$_x$Y$_2$Te$_5$ with different Pd content. Impurity phases, as marked by the red arrows, appear only for samples with high doping level. **d**, Lattice parameter b as a function of x in Pd$_x$HoTe$_3$ and Pd$_x$Y$_2$Te$_5$. The straight line guides the eyes.

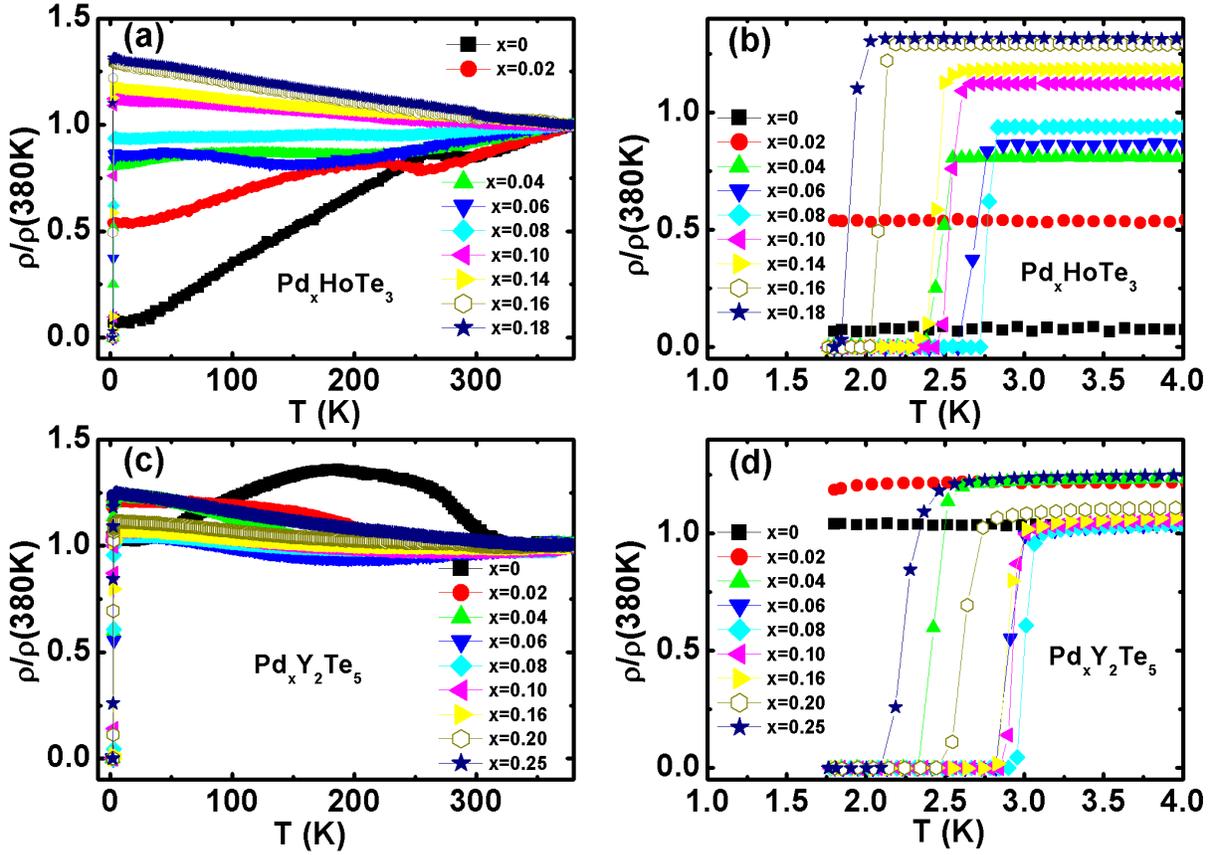

**Figure 2. Transport properties of $Pd_xHoTe_3$ and $Pd_xY_2Te_5$ with different Pd content.** **a**, **c**, Normalized temperature dependent resistivity data measured on $Pd_xHoTe_3$ (0≤x≤0.18) and $Pd_xY_2Te_5$ (0≤x≤0.25), respectively. With increasing x, the hump is suppressed and superconductivity emerges. **b**, **d**, *T*-dependent resistivity in an expanded region for $Pd_xHoTe_3$ and $Pd_xY_2Te_5$ with different Pd content. Superconductivity appears for x=0.04, and $T_c$ reaches a maximum for x=0.08. Further increasing Pd content, $T_c$ is slowly suppressed.

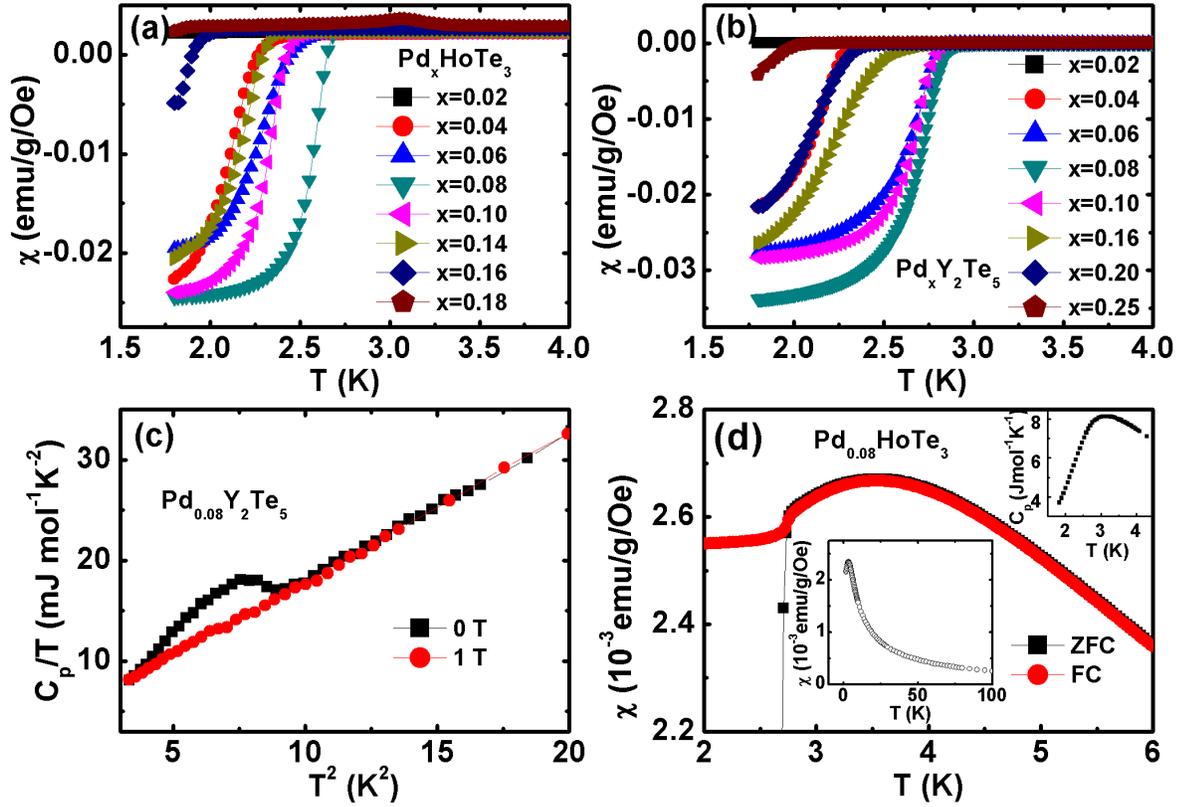

**Figure 3. Characterization of the superconducting phase transitions for $Pd_xHoTe_3$ and $Pd_xY_2Te_5$ with different Pd content. a**, Low temperature magnetic susceptibility $\chi(T)$ for $Pd_xHoTe_3$ with x= 0.02-0.18. Data was collected in zero field cooled (ZFC) mode. **b**, $\chi(T)$ vs. T for $Pd_xY_2Te_5$ with x= 0.02-0.25. **c**, The plot of C/T vs. T for the optimal doping sample $Pd_{0.08}Y_2Te_5$ at $H$=0 and 1T. **d**, $\chi(T)$ vs. $T$ for $Pd_{0.08}HoTe_3$ in the low temperature region with a magnetic field of 10 Oe. Upper inset: Heat capacity $C_p$ vs. $T$ for $Pd_{0.08}HoTe_3$. Lower inset: Temperature dependence of magnetic susceptibility $\chi(T)$ for $Pd_{0.08}HoTe_3$ measured with $H$=0.1 T.

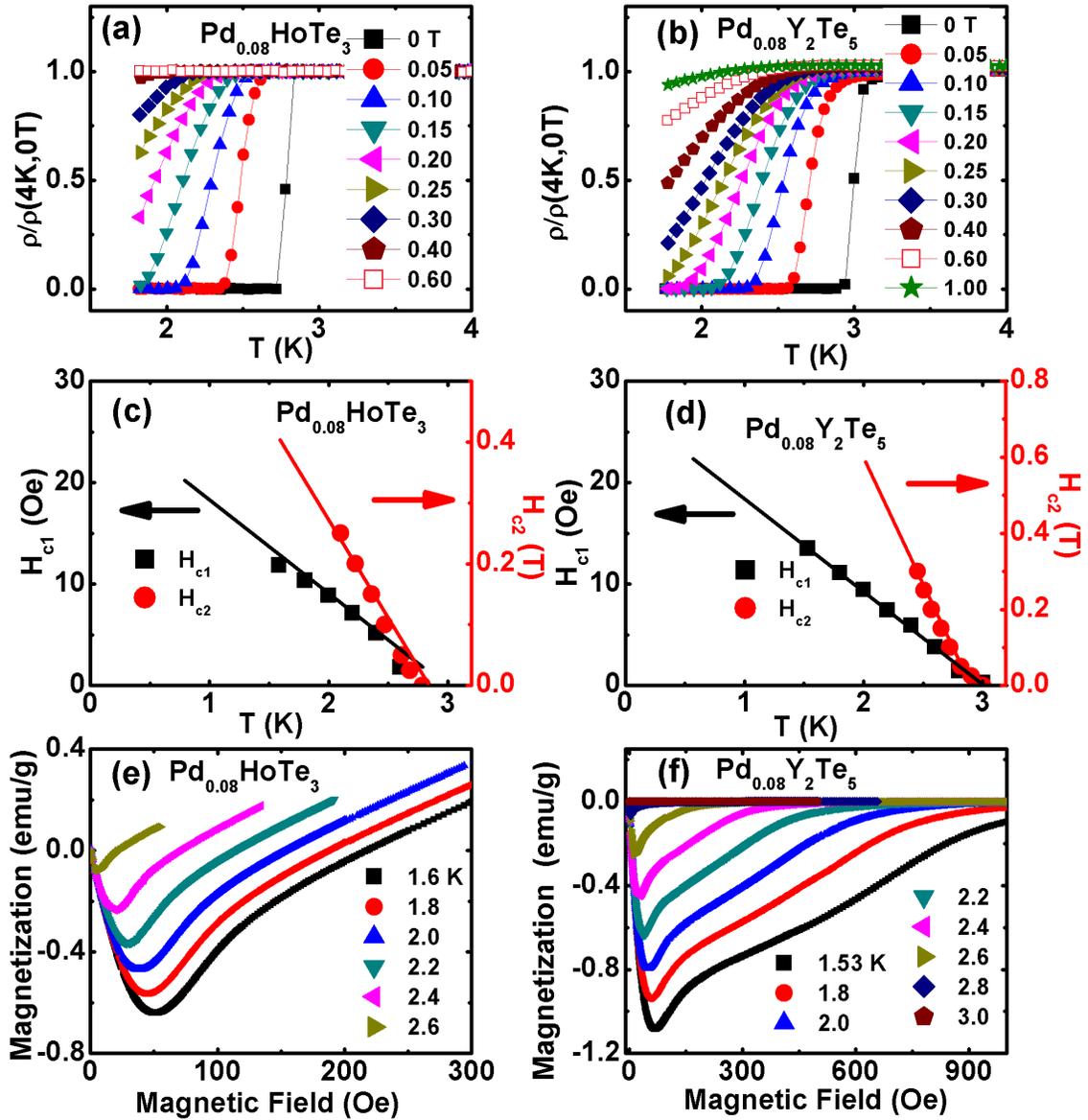

**Figure 4. Characterization of the superconductivity state in Pd$_{0.08}$HoTe$_3$ and Pd$_{0.08}$Y$_2$Te$_5$. a, b,** Temperature dependence of electrical resistivity for optimal doped Pd$_{0.08}$HoTe$_3$ and Pd$_{0.08}$Y$_2$Te$_5$ under various magnetic fields. **c, d,** $H_c$-$T$ phase diagram, $H_{c1}$ and $H_{c2}$, determined from $M(H)$ and $\rho(H)$ data. **e, f,** M versus H plots for selected temperatures at low magnetic fields.

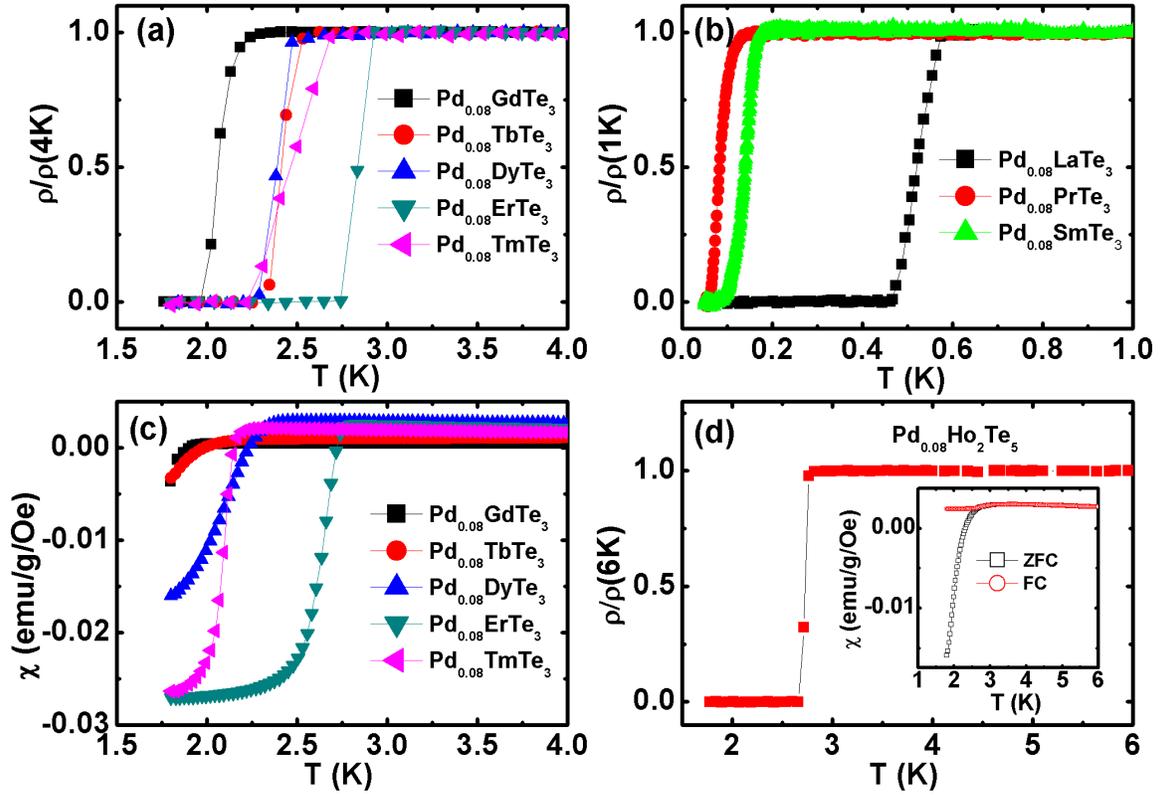

**Figure 5. Characterization of the superconductivity transitions for $Pd_{0.08}R_ETe_n$. a,** Normalized temperature dependent resistivity for heavy rare earth tri-tellurides $Pd_{0.08}GdTe_3$, $Pd_{0.08}TbTe_3$, $Pd_{0.08}DyTe_3$, $Pd_{0.08}ErTe_3$, and $Pd_{0.08}TmTe_3$. The critical temperatures $T_c$ are 2.3, 2.55, 2.5, 2.9, 2.7 K, respectively. **b,** Low temperature resistivity data for light rare earth tri-tellurides $Pd_{0.08}LaTe_3$, $Pd_{0.08}PrTe_3$, and $Pd_{0.08}SmTe_3$, which superconduct at 0.6, 0.13, and 0.19 K, respectively. **c,** Temperature dependent dc magnetization in heavy rare earth tri-tellurides (Gd-Tm). **d,** Temperature dependent resistivity and dc magnetization for $Pd_{0.08}Ho_2Te_5$ ($T_c$=2.8 K).

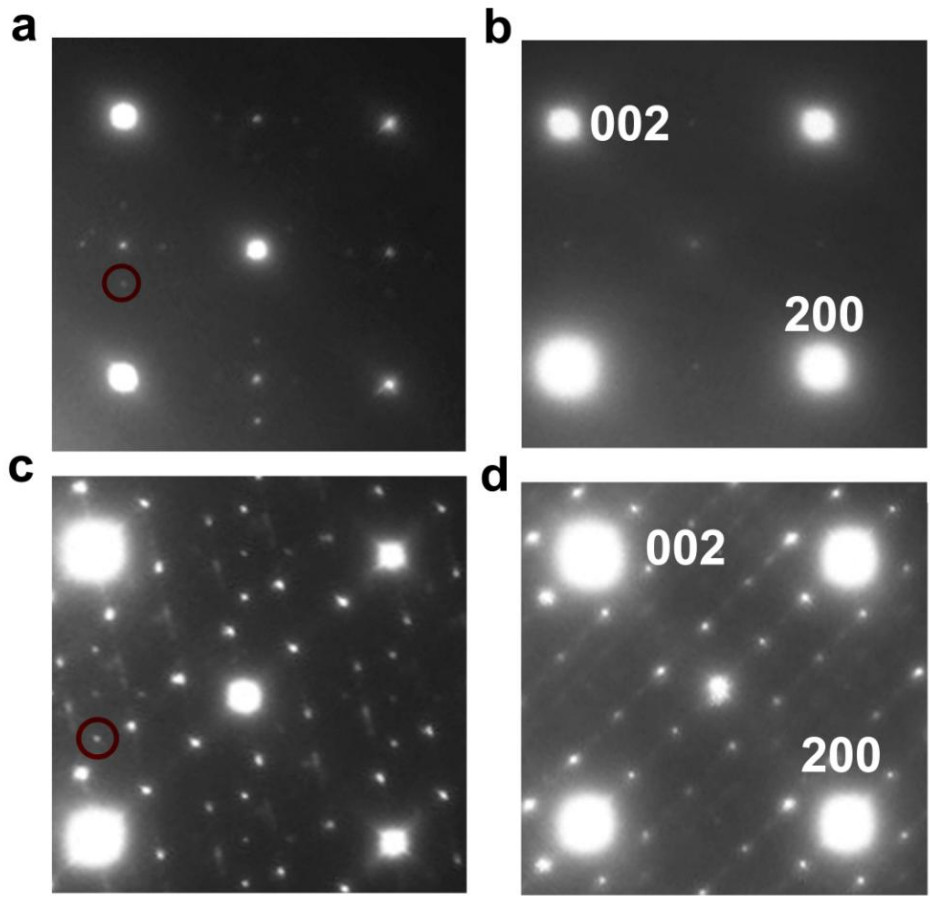

**Figure 6 Structural modulations in parent CDW phases and superconducting phases.** Electron diffraction patterns taken along the [010] zone-axis direction obtained at about 100 K. (a) $TmTe_3$, (c) $Y_2Te_5$ and (b) $Pd_{0.08}TmTe_3$, (d) $Pd_{0.08}Y_2Te_5$. Structural modulation from incommensurate CDW wave in the double Te planes can be clearly seen in the parent phases as indicated by circles.

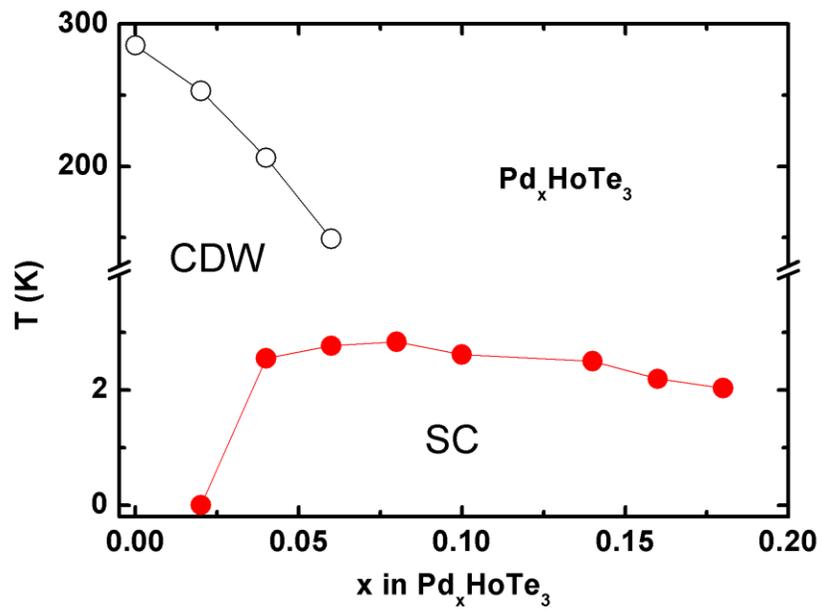

**Figure 7. Electronic phase diagram for Pd$_x$HoTe$_3$.** Open and closed circles are determined from resistivity, which correspond to the first CDW transition temperature and superconducting transition temperature, respectively.